# Microwave spectroscopy on a quantum-dot molecule

T.H. Oosterkamp[1], T. Fujisawa[1,2], W.G. van der Wiel[1], K. Ishibashi[1,3],

R.V. Hijman[1], S. Tarucha[2] and L.P. Kouwenhoven[1].

[1] Department of Applied Physics and DIMES, Delft University of Technology,
P.O. Box 5046, 2600 GA Delft, The Netherlands.
[2] NTT Basic Research Laboratories, 3-1, Morinosoto Wakamiya, Atsugi-shi,
Kanagawa 243-0198, Japan.
[3] Institute of Physical and Chemical Research (RIKEN), 2-1, Hirosawa,
Wako, Saitama 351-01, Japan.

Quantum dots are small conductive regions in a semiconductor containing a variable number of electrons (N=1 to 1000) that occupy well defined discrete quantum states. They are often referred to as artificial atoms[1] with the unique property that they can be connected to current and voltage contacts. This allows one to use transport measurements to probe the discrete energy spectra. To continue the analogy with atoms two quantum dots can be connected to form an 'artificial molecule'. Depending on the strength of the inter-dot coupling the two dots can have an ionic binding[2-6] (i.e. electrons are localized on the individual dots) or a covalent binding (i.e. electrons are delocalized over both dots). The covalent binding leads to a bonding and an antibonding state with an energy splitting proportional to the tunnel coupling. In the dc current response to microwave excitation[5-8] we observe a transition from an ionic bonding to a covalent bonding when we vary the inter-dot coupling strength. This demonstrates controllable quantum coherence in single electron devices.

When particles are allowed to tunnel back and forth between two quantum systems, the energy states of the individual systems mix and form new states that extend over both systems. The extended states are referred to as the bonding- or symmetric state, and the anti-bonding- or anti-symmetric state. In solid state systems the energy splitting between bonding and anti-bonding states have been observed in quantum well structures[9,10], superconducting tunneling devices[11,12], and exciton systems[13].

Quantum dots are uniquely engineered solid state systems in the sense that they have discrete states and the electrons on the dots are strongly interacting. The question whether different dots can be coupled together in a quantum-mechanically coherent way is non-trivial. The reason is that quantum dots composing single-electron devices are embedded in an environment with many electronic degrees of freedom. The electron that occupies the covalent state of a double dot system, has a Coulomb interaction with all the other electrons confined on the dots and also with the electrons in the current and voltage leads. These interactions can lead to dephasing of the quantum mechanical wave function resulting in a breakdown of the covalent state. For realistic devices there is yet no theory that can calculate reliable dephasing rates.

Nevertheless, if elements like quantum dots will ever be integrated in little quantum circuits[14-16], it is necessary that dots can be coupled coherently. This letter reports on experiments that demonstrate this ability and, in our opinion, the cleanliness of the results are promising for manipulation of electrons in more complicated circuits. We have used microwave spectroscopy (0-50 GHz) to measure the energy differences between states in the two dots of the device[2] shown in Fig. 1a. We show that these energy differences, including the bonding-antibonding splitting, is controlled by gate voltages which tune the tunnel coupling between the dots. We first discuss the weak-coupling regime.

Electrons are strongly localized on the individual dots when tunneling between the two dots is weak. Electron transport is then governed by single-electron charging effects[8]. The charging energies can be tuned away by means of the gate voltages. It is then energetically allowed for an electron to tunnel between dots when a discrete state in the left dot is aligned with a discrete state in the right dot. External voltages also control the alignment of the discrete states. A current can flow when electrons can tunnel, while conserving energy, from the left lead, through the left and right dots, to the right lead. Note that energy is also conserved when photons of energy $hf$ are absorbed from the microwave field which match the energy difference between the states of the two dots (see Fig. 1b).

The resonance in the lowest trace in Fig. 1c is due to an alignment of discrete states. The other traces are measured while applying a microwave signal. The satellite resonances are due to photon assisted tunneling processes which involve the emission (left satellite) or absorption (right satellite) of a microwave photon.

Stoof and Nazarov[17] give a detailed description of photon assisted tunneling in a double quantum dot. The basic idea is that electrons can absorb fixed quanta of energy $hf$ from a classical oscillating field. An ac voltage drop $V = V_{ac}\cos(2\pi ft)$ across a tunnel barrier modifies the tunnel rate through the barrier as[18]:

$$\widetilde{\Gamma}(E) = \sum_{n=-\infty}^{+\infty} J_n^2(\alpha)\Gamma(E + nhf). \quad (1)$$

Here $\widetilde{\Gamma}(E)$ and $\Gamma(E)$ are the tunnel rates at energy $E$ with and without an ac voltage, respectively. $J_n^2(\alpha)$ is the square of the $n^{th}$ order Bessel function evaluated at $\alpha = \frac{eV_{ac}}{hf}$, which describes the probability that an electron absorbs or emits $n$ photons of energy $hf$.

Figure 2 shows the current for several microwave powers. The dashed curve shows the main resonance measured at zero power. As the power is increased, satellite peaks appear corresponding to the absorption of multiple photons which are observed up to $n = 11$. At these high powers the microwaves strongly perturb tunneling. This is reflected by the non-linear dependence of the peak heights on power (left inset of Fig. 2), which is in agreement with the expected Bessel function behavior.

The right inset to Fig. 2 shows that the separation of the satellite peaks from the main peak depends linearly on frequency between 1 and 50 GHz. As we discuss below, this linearity implies that the tunnel coupling is negligible. The electrons are thus localized on the individual dots and they have an ionic bonding. The line proportional to $2hf$ is taken from data at higher microwave powers where electrons absorb or emit two photons during tunneling.

In contrast to the case of weakly-coupled dots, covalent bonding occurs when two discrete states that are spatially separated become *strongly* coupled. Electrons then tunnel quickly back and forth between the dots. In a quantum mechanical description this results in a bonding and an anti-bonding state which are lower and higher in energy, respectively, than the original states. Our strong-coupling measurements were made on a second type of double-dot sample (see inset to Fig. 4). To single out the current only due to microwaves we operate the device as an electron pump driven by photons in a way described theoretically by Stafford and Wingreen[19] and by Brune et al.[20] (see the diagrams of Fig. 3a-c). By sweeping the gate voltages we vary $\Delta E = E_{left} - E_{right}$, where $E_{left}$ and $E_{right}$ are the energies of the uncoupled states in the left and right dot. The bonding and anti-bonding states, that are a superposition of the wavefunctions corresponding to an electron in the left or in the right dot, have an energy splitting of $\Delta E^* = E_{anti-bond} - E_{bond} = \sqrt{(\Delta E)^2 + (2T)^2}$, where $T$ is the tunnel coupling between the two dots. When the sample is irradiated, a photo current may result as illustrated in Fig. 3a-c. A non-zero current indicates that an electron was excited from the bonding state to the anti-bonding state, thereby fulfilling the condition $hf = \Delta E^*$, or conversely

$$\Delta E = \sqrt{(hf)^2 - (2T)^2}. \quad (2)$$

Figure 3 shows measured current traces as a function of the uncoupled energy splitting $\Delta E$, where from top to bottom the applied microwave frequency is decreased from 17 to 7.5 GHz in 0.5 GHz steps. The distance between the pumping peaks, which is proportional to $2\Delta E$, decreases as the frequency is lowered. However, the peak distance decreases faster than linear in frequency; the peaks follows the hyperbola rather than the straight lines. The distance goes to zero when the frequency approaches the minimum energy gap between bonding and anti-bonding states, $hf = 2T$. For frequencies smaller than the coupling, $hf < 2T$, the photon energy is too small to induce a transition from the bonding to the anti-bonding state.

The coupling between the dots can be decreased by changing the gate voltage on the center gate to more negative values or by applying a magnetic field perpendicular to the sample. In Fig. 4 we have plotted the energy spacing $\Delta E$ at which the pumping current is at a maximum, as a function of frequency. Different labels correspond to different center gate voltage settings and magnetic fields. The solid lines are fits of equation (2) to the measured data. It follows that the coupling $2T$ has been tuned from 11 to 60 $\mu$eV. The good agreement with equation 2 and the clear non-linear frequency dependence demonstrates the control over the formation of a covalent bonding between the two dots.

Quantum dots have been suggested as possible candidates for building a quantum computer[14–16]. We have shown that it is indeed possible to coherently couple dots, and that one can induce transitions between the extended states. The next crucial step towards quantum logic gates is to show that the coherence of the superposition is preserved on time scales much longer than the time needed for manipulating the electron wave functions. A lower bound for the dephasing time is $\tau_\varphi > 1$ ns, which we deduce from our narrowest peaks and from the smallest energy gaps between the bonding and anti-bonding states that we have resolved. Future experiments include measurements of the decoherence time in which the states are manipulated by applying the microwaves in short pulses.

**Acknowledgments.** We thank R. Aguado, S.M. Cronenwett, S.F. Godijn, P. Hadley, C.J.P.M. Harmans, K.K. Likharev, J.E. Mooij, Yu. V. Nazarov, R.M. Schouten, T.H. Stoof and N.C. van der Vaart for experimental help and useful discussions. This work was supported by the Dutch Organization for Research on Matter (FOM) and by the EU via the TMR network. L.P.K. was supported by the Dutch Royal Academy of Arts and Sciences (KNAW).

**Figure captions:**

Figure 1a) Photo of the double quantum dot sample. The source and drain regions as well as the left and right dots are indicated schematically. The tunnel barriers are depicted as arrows. The metallic gates (1, 2, 3, and F) are fabricated on top of a GaAs/AlGaAs heterostructure with a 2 dimensional electron gas (2DEG) 100 nm below the surface. At 4.2 K the 2DEG mobility is $2.3 \times 10^6$ cm$^2$/Vs and the electron density is $1.9 \times 10^{15}$ m$^{-2}$. Applying negative voltages to all the gates depletes the electron gas underneath them and forms two dots with estimated sizes of $(170\,\text{nm})^2$ and $(130\,\text{nm})^2$. We measure the dc photo current in response to a microwave signal (0-50 GHz) that is capacitatively coupled to gate 2. The tunnel coupling between the two dots and to the reservoirs can be controlled with the voltages on gate 1, 2 and 3. The dots contain about 60 and 35 electrons, respectively. The sample is cooled in a dilution refrigerator yielding an electron temperature in the source and drain contacts of ∼100 mK. b) Diagram of the electron energies in the dot for the case that an electron needs to absorb a photon in order to contribute to the current. Shaded areas represent the electron states in the leads that are continuously filled up to the Fermi levels. A voltage $V_{SD}$ applied between the source and drain contacts shifts one Fermi level relative to the other. The discrete energy states in the two dots can be adjusted independently by changing the gate voltages. c) The upper diagrams illustrate three situations of the energy state in the left dot relative to the state in the right dot. The hatched lines denote the Fermi levels in the leads. The bottom curve shows the current as a function of the voltage on gate 1 for $V_{SD} = 500$ $\mu$V without applying microwaves. A single resonance occurs when two states line up. Other curves, which have been offset for clarity, show the current when microwaves with frequency $f$ from 4 to 10 GHz are applied. Now, two additional satellite resonances occur when the two states are exactly a photon energy apart. The corresponding photon-assisted tunneling processes are illustrated in the upper diagrams.

Figure 2. Current versus gate voltage of a weakly coupled double-dot. The dashed curve is without microwaves and only contains the main resonance. The solid curves are taken at 8 GHz for increasing microwave powers resulting in an increasing number of satellite peaks. At the right side of the main peak these correspond to photon absorption. The source drain voltage $V_{SD} = 700$ $\mu$V and the photon energy $hf = 32$ $\mu$eV at 8 GHz. At the highest power we observe eleven satellite peaks demonstrating multiple photon absorption. Left inset: Height of the first four satellite peaks as a function of the microwave amplitude. The observed height dependence agrees with the expected Bessel function behavior. Right inset: Distance between main resonance and first two satellites as a function of the applied frequency from 1 to 50 GHz. The distance is transferred to energy through $\Delta E = \kappa \Delta V_g$ where $\kappa$ is the appropriate capacitance ratio for our device that converts gate voltage to energy[8]. The agreement between data points and the two solid lines, which have slopes of $h$ and $2h$, demonstrates that we observe the expected linear frequency dependence of the one and two photon processes.

Figure 3a) - c) Energy diagrams. Solid lines depict the energy states $E_{left}$ and $E_{right}$ in the two dots for the case that the coupling is weak and that their energy difference is simply $\Delta E = E_{left} - E_{right}$. When the dots are strongly-coupled, the states delocalize over both dots thereby forming a bonding and an anti-bonding state. These are indicated by two dotted lines. Their energy difference is $\Delta E^* = \sqrt{\Delta E^2 + (2T)^2}$. Electrons are transferred from the bonding to the anti-bonding state when $\Delta E^* = hf$. In (a) $E_{left} > E_{right}$ which results in electron pumping from right to left corresponding to a negative current. In (b) the whole system is symmetric ($E_{left} = E_{right}$) and consequently the net electron flow must be zero. In (c) $E_{left} > E_{right}$ which gives rise to pumping from left to right and a positive current. (d) Measured pumped current through the strongly-coupled double-dot. Gates 1 and 3 are swept simultaneously in such a way that we vary the energy difference $\Delta E$. The different traces are taken at different microwave frequencies and are offset such that the right vertical axis gives the frequency. The main resonance is absent since we have set $V_{SD} = 0$. The satellite peaks typically have an amplitude of 0.5 pA. For weakly-coupled dots the satellite peaks are expected to move linearly with frequency, thereby following the straight dashed lines. In contrast, we observe that the satellite peaks follow the fitted hyperbola $hf = \sqrt{\Delta E^2 + (2T)^2}$ using $T$ as a fitting parameter.

Figure 4. Half the spacing in gate voltage between the positive and negative satellite peaks as a function of frequency. Gate voltage spacing has been transferred to energy difference $\Delta E$ (see also figure caption 2). Different curves correspond to different coupling constants $T$. Solid lines are theoretical fits to $\Delta E = \sqrt{(hf)^2 - (2T)^2}$. The resulting values for $2T$ are given in the figure. In the limit of weak-coupling this reduces to $\Delta E = hf$ which is indicated by the dashed line. The coupling is varied by applying different voltages to the center gate (2) or by changing the magnetic field ($\blacklozenge$ : $B = 3.3$ T; $\blacksquare$ : $B = 2.2$ T; other curves: $B = 0$). The upper left inset shows a diagram of the sample[5]. A narrow channel is defined by locally depleting the 2DEG using focussed ion beam implantation (FIB). Two dots are then formed by applying negative voltages to the three gates (1, 2, 3) that cross the channel. Microwaves are capacitively coupled to gate 2.

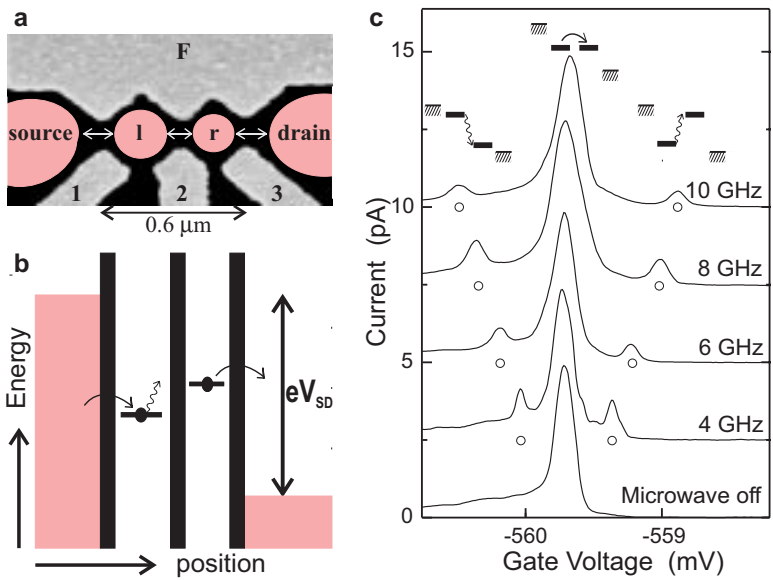

Figure 1

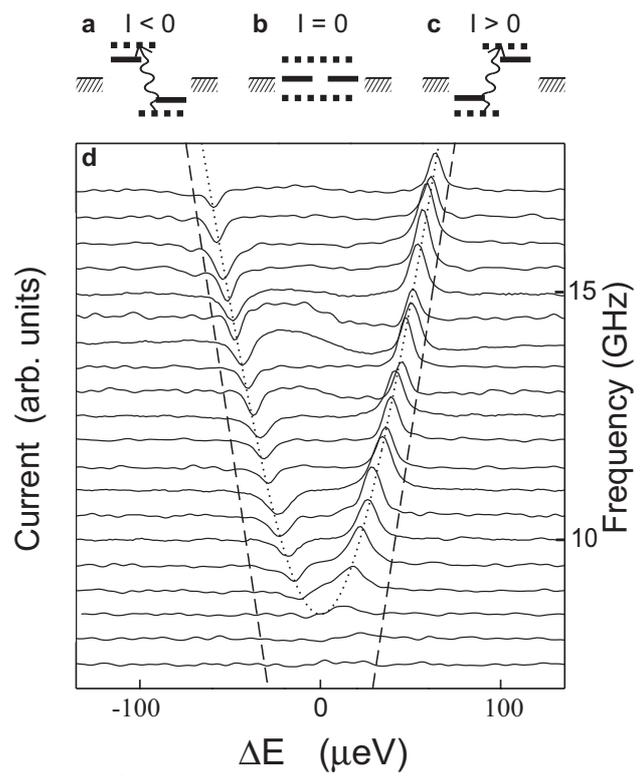

Figure 3

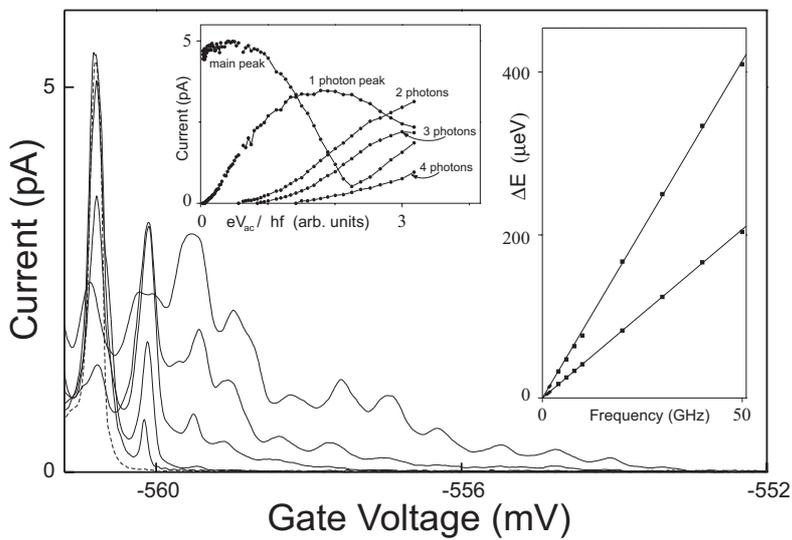

Figure 2

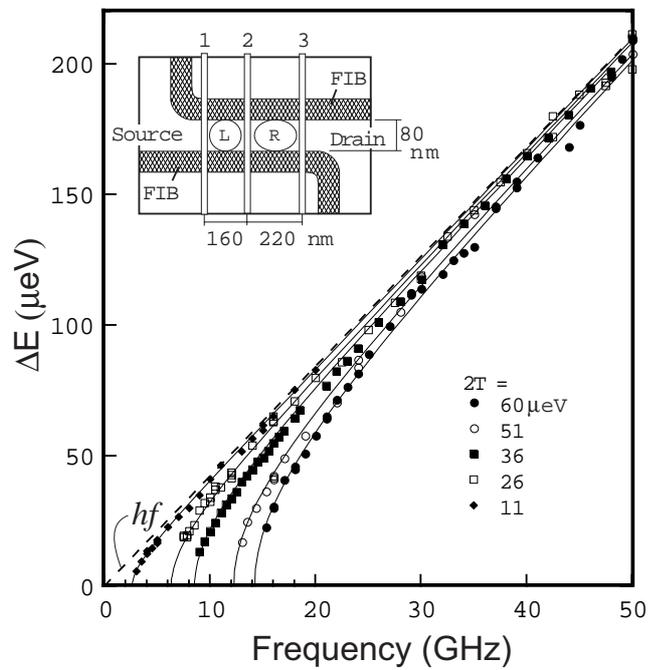

Figure 4
Oosterkamp et al.